\documentclass[review,number,sort&compress]{elsarticle}

\usepackage{graphicx}
\usepackage[squaren]{SIunits}

\newcommand{\gfrac}[2]{\displaystyle\frac{#1}{#2}}

\newcommand{\dd}{\mbox{d}}

\journal{Nucl.\ Instrum.\ Meth.\ A}

\begin{document}

\begin{frontmatter}

\title{Compton polarimetry revisited}

\author[add1]{D.~Bernard\corref{cor}}
\ead{denis.bernard at in2p3.fr}

\address[add1]{LLR, Ecole Polytechnique, CNRS/IN2P3,
91128 Palaiseau, France}


\begin{abstract}
We compute the average polarisation asymmetry from the Klein-Nishina
differential cross section on free electrons at rest.
As expected from the expression for the asymmetry, the average
asymmetry is found to decrease like the inverse of the incident photon
energy asymptotically at high energy.
We then compute a simple estimator of the polarisation fraction that
makes optimal use of all the kinematic information present in an event
final state, by the use of ``moments'' method, and we compare its
statistical power to that of a simple fit of the azimuthal
distribution. 
In contrast to polarimetry with pair creation, for which we obtained an
improvement by a factor of larger than two in a previous work, here for Compton
scattering the improvement is only of 10--20\,\%.
\end{abstract}

\end{frontmatter}

\section{Cosmic-source polarimetry: the high-energy frontier}

Polarimetry is a powerful diagnostic of specific phenomena at work in
cosmic sources in the radio-wave and optical energy bands, but very
few results are available at high photon energies: 
the only significant observation in the X-gamma energy range, to date, is the
measurement of a linear polarisation fraction of $P = 19 \pm 1 \%$ of
the 2.6\,keV emission of the Crab nebula by a Bragg polarimeter on
board OSO-8 \cite{Weisskopf:1978}.
At higher energies, hard-X-ray and soft-gamma-ray telescopes that have
flown to space in the past (COMPTEL \cite{Lei:1997},
BATSE\cite{Willis:2005bp}) were not optimized for polarimetry, and
their sensitivity to polarisation was poor.
Presently active missions 
(Integral IBIS\cite{Forot:2008ud,Laurent:2003} and
SPI \cite{Chauvin:2013aka})
have provided some improvement, with, in particular, mildly significant
measurements of
$P = 28 \pm 6 \%$ (130 to 440 keV \cite{Chauvin:2013aka})
and
$P=47_{-13}^{+19}\%$ (200 to 800 keV \cite{Forot:2008ud})
for the Crab Nebula.
A number of Compton polarimeter/telescope projects have been
developed, some of which also propose to record photon conversions to
$e^+ e^-$ pairs.
A variety of technologies have been considered, such as 
scintillator arrays
(POGO \cite{POGO}, GRAPE \cite{GRAPE}, POLAR \cite{POLAR}),
Si or Ge microstrip detectors 
(MEGA \cite{MEGA}, ASTROGAM \cite{ASTROGAM:2015})
or combinations of these (Si $+$ LaBr$_3$ for GRIPS \cite{GRIPS},
Si $+$ CsI(Tl) for TIGRE \cite{TIGRE}),
semiconductor pixel detectors (CIPHER \cite{CIPHER})
and
liquid xenon (LXeGRIT \cite{LXeGRIT}) time projection chambers (TPC).
In most Compton telescopes the reconstruction of the direction of the
incident photon provides an uncertainty area which has the shape of a
thin cone arc. 
The tracking of the recoil electron from the first Compton interaction
with a measurement of the direction of the recoil momentum, as is
within reach with a gas TPC, allows to decrease the length of the arc
and therefore to improve dramatically the sensitivity of the detector
(\cite{Tanimori:2015wma} and references therein).

Some of these telescopes are sensitive to photon energies up to tens
of MeV in the Compton mode, but their sensitivity to polarisation
above a few MeV is either nonexistent 
or undocumented.

\section{Polarisation asymmetry and average polarisation asymmetry}

As is well known, the sensitivity to polarisation of Compton
scattering is excellent at low energies (Thomson scattering), as the
polarisation asymmetry $\cal A$, 
 also known as the modulation factor and
defined by the phase-space dependence of the differential cross section, 
\begin{equation}
\gfrac{\dd \sigma}{\dd \Omega}
\propto
\left[ 1 + {\cal A} P \cos{[2(\phi-\phi_0)]} \right],
\label{eq:def:diff:cross:section:compton}
\end{equation}
reaches $-1$ at a polar angle $\theta$ of $90^\circ$ (Fig. 1 of Ref. \cite{Krawczynski:2011fm}).
In this expression, 
$\phi$ is the azimuthal angle, that is the angle between the
scattering plane and the direction of polarisation of the incident
photon.
Unfortunately, $\cal A$ is decreasing with energy, and as the
precision of the measurement scales as 
$\sigma_P \propto 1/({\cal A} \sqrt{N})$
when the background noise is negligible and where $N$ is
the number of signal event, the sensitivity of Compton polarimetry
decreases at high energies.
With the goal of a quantitative assessment of this sensitivity, in this
paper we compute the average polarisation asymmetry 
$\langle \cal A \rangle$ from the Klein-Nishina differential cross
section on free electrons at rest
\cite{Klein:Nishina:1929,Nishina:1929}.
$\langle \cal A \rangle$ is defined from the differential cross
section in $\phi$, that is after the full differential cross section
(eq. (\ref{eq:def:diff:cross:section:compton}))
has been integrated over the other variables that describe the final
state:
\begin{equation}
\gfrac{\dd \sigma}{\dd \phi}
\propto
\left[ 1 + \langle {\cal A} \rangle P \cos{[2(\phi-\phi_0)]} \right].
\label{eq:def:diff:cross:section:compton:int}
\end{equation}
Following Heitler \cite{Heitler:1954}, the doubly differential cross
section for linear polarised radiation reads:
\begin{equation}
\gfrac{\dd \sigma}{\dd \Omega}
=
\gfrac{r_0^2}{2} x^2 
\left[
x + \gfrac{1}{x} - 2 \sin^2{\theta}\cos^2{\phi}
\right],
\label{eq:diff:cross:section:compton}
\end{equation}
where $x=k/k_0$, $k_0$ and $k$ are the energy of the incident and
scattered $\gamma$s, respectively;
$\theta$ is the scattering angle, that is the polar angle of the
direction of the scattered $\gamma$ with respect to the direction of
the incident $\gamma$.
The differential element $\dd \Omega$ is $\sin \theta \dd
\theta\dd\phi$ as usual.
In the case of partially polarised emission with polarisation
fraction $P$, the differential cross section becomes:
\begin{equation}
\dd \sigma = \gfrac{r_0^2}{2} x^2 \left[
x + \gfrac{1}{x} - \sin^2{\theta} \left(\cos{(2\phi)} P + 1 \right) \right]
\sin\theta \dd \theta \dd \phi.
\end{equation}

\begin{figure} [ht]
 \includegraphics[width=\linewidth]{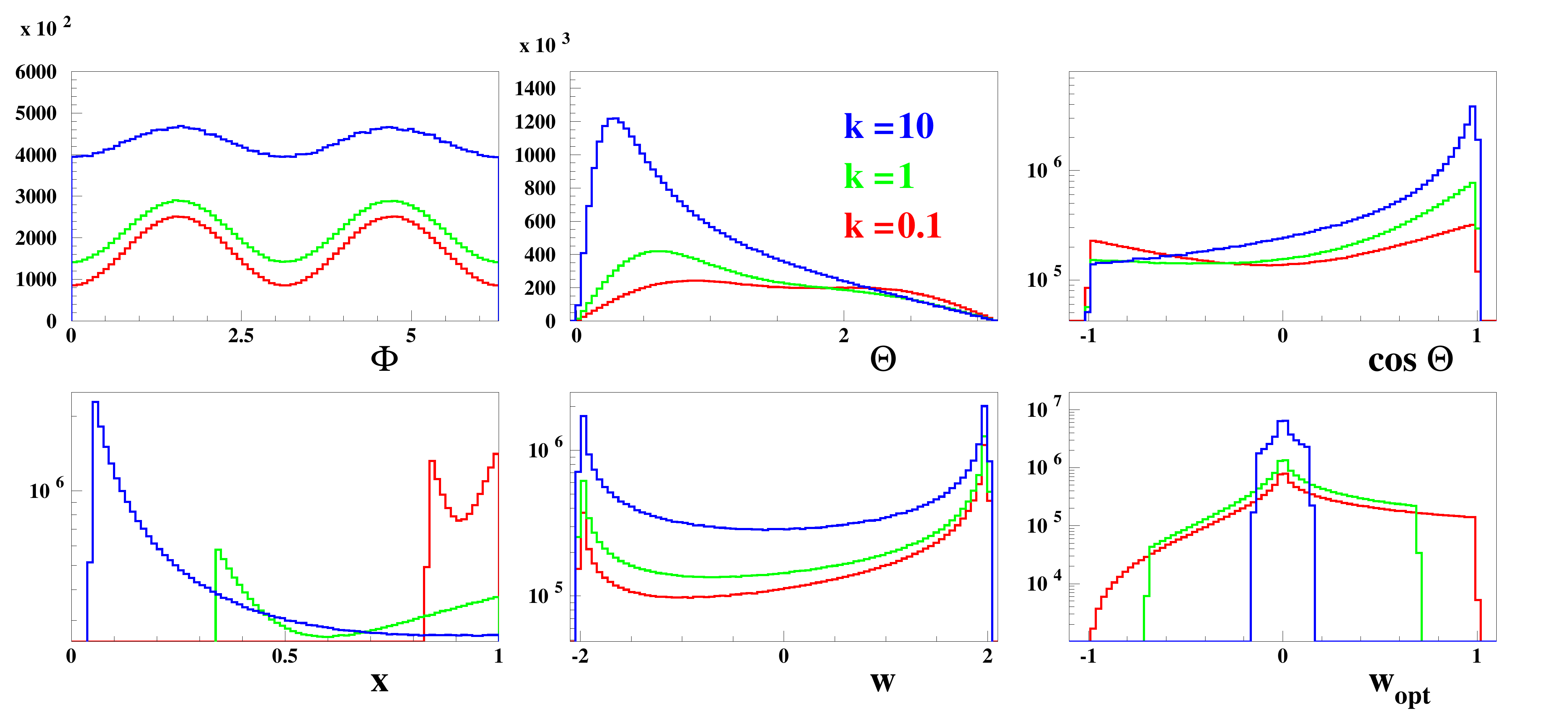}
 \caption{\label{fig:compton:spectra} 
Spectra of the azimuthal and polar angles $\phi$ and $\theta$, of
$\cos\theta$, of the fraction $x$ of the incident photon energy
carried away by the scattered photon, and of the 1D and 2D weights
$w$ and $w_{\text{opt}}$,
for incident photon energies $0.1 mc^2$, $mc^2$,
and $10 mc^2$, all for a fully polarized beam. }
\end{figure}

The minus sign reflects the fact that photons Compton scatter
preferentially into the direction perpendicular to the orientation of
the electric field of the incoming radiation.
The energy of the scattered $\gamma$ is related to
$\theta$ from energy-momentum conservation:
$ k = k_0/[1+k_0(1 - \cos\theta)] $, 
~
$ x = 1/[1+k_0(1 - \cos\theta)] $, 
~
$ \cos\theta = 1 - \left({1}/{x} -1 \right)/k_0 $,
~
$ \sin\theta \dd \theta = - \dd x /(k_0 x^2)$
~
and
~
$ \sin^2\theta = 2 \left[1/(x k_0) - 1/k_0 \right] - \left[1/(x k_0) - 1/k_0 \right]^2 $.
We then obtain \cite{Heitler:1954,QED:Akhiezer-Berestetskii}:
\begin{equation}
\dd \sigma
= \gfrac{r_0^2}{2 k_0} \left[
x + \gfrac{1}{x} - 
\left[
 2
\left(\gfrac{1}{x k_0} - \gfrac{1}{k_0} \right)
-
\left(\gfrac{1}{x k_0} - \gfrac{1}{k_0} \right)^2
\right]
\left(\cos{(2\phi)} P + 1 \right) 
\right]
 \dd x \dd \phi.
\end{equation}
$k$ varies in a range such that $-1 \le \cos\theta \le 1$, that is 
$ 1/(1 + 2 k_0) \le x \le 1 $.
The distributions of these kinematic variables are shown in
Fig. \ref{fig:compton:spectra}.
After an elementary integration over $x$, we obtain:
\begin{eqnarray}
\dd \sigma
& = & r_0^2
\left[
\frac{1+k_0}{k_0^3}
\left(
\frac{2 k_0 (1+k_0)}{1+2 k_0} - \ln{(1+2 k_0)} 
\right)
+ \frac{\ln(1+2 k_0)}{2 k_0}
\right.
\label{eq:phi:diff:cross:section}
\\
 & &
\left.
- \frac{1+3 k_0}{(1+2 k_0)^2}
+ \frac{\left(2 k_0-(k_0+1) \log(2 k_0+1)\right)}{k_0^3}
 \cos{(2\phi)}P
\right]
 \dd \phi, \nonumber
\end{eqnarray}

that is a total cross section of
\cite{Klein:Nishina:1929,Heitler:1954,QED:Akhiezer-Berestetskii}:
\begin{eqnarray}
\sigma
& = & 2 \pi r_0^2 \times 
\\
& &
\left[
\frac{1+k_0}{k_0^3}
\left( \frac{2 k_0 (1+k_0)}{1+2 k_0} - \ln{(1+2 k_0)} \right)
+ \frac{\ln(1+2 k_0)}{2 k_0}
- \frac{1+3 k_0}{(1+2 k_0)^2}
\right].
\nonumber
\end{eqnarray}

Equating the constant term and the term proportional to
$\cos{(2\phi)}P$ in eqs. (\ref{eq:def:diff:cross:section:compton:int})
and (\ref{eq:phi:diff:cross:section}), we obtain for the average
polarisation asymmetry:
\begin{eqnarray}
\langle {\cal A} \rangle = 
\frac{(2 k_0-(k_0+1) \log(2 k_0+1))}
{
(1+k_0) \left( \gfrac{2 k_0 (1+k_0)}{1+2 k_0} - \ln{(1+2 k_0)} \right)
+ \gfrac{k_0^2\ln(1+2 k_0)}{2}
- \gfrac{(1+3 k_0)k_0^3}{(1+2 k_0)^2}
}.
\label{eq:compton:asymmetry:full}
\end{eqnarray}

\begin{figure} [ht]
 \begin{center}
 \includegraphics[width=0.7\linewidth]{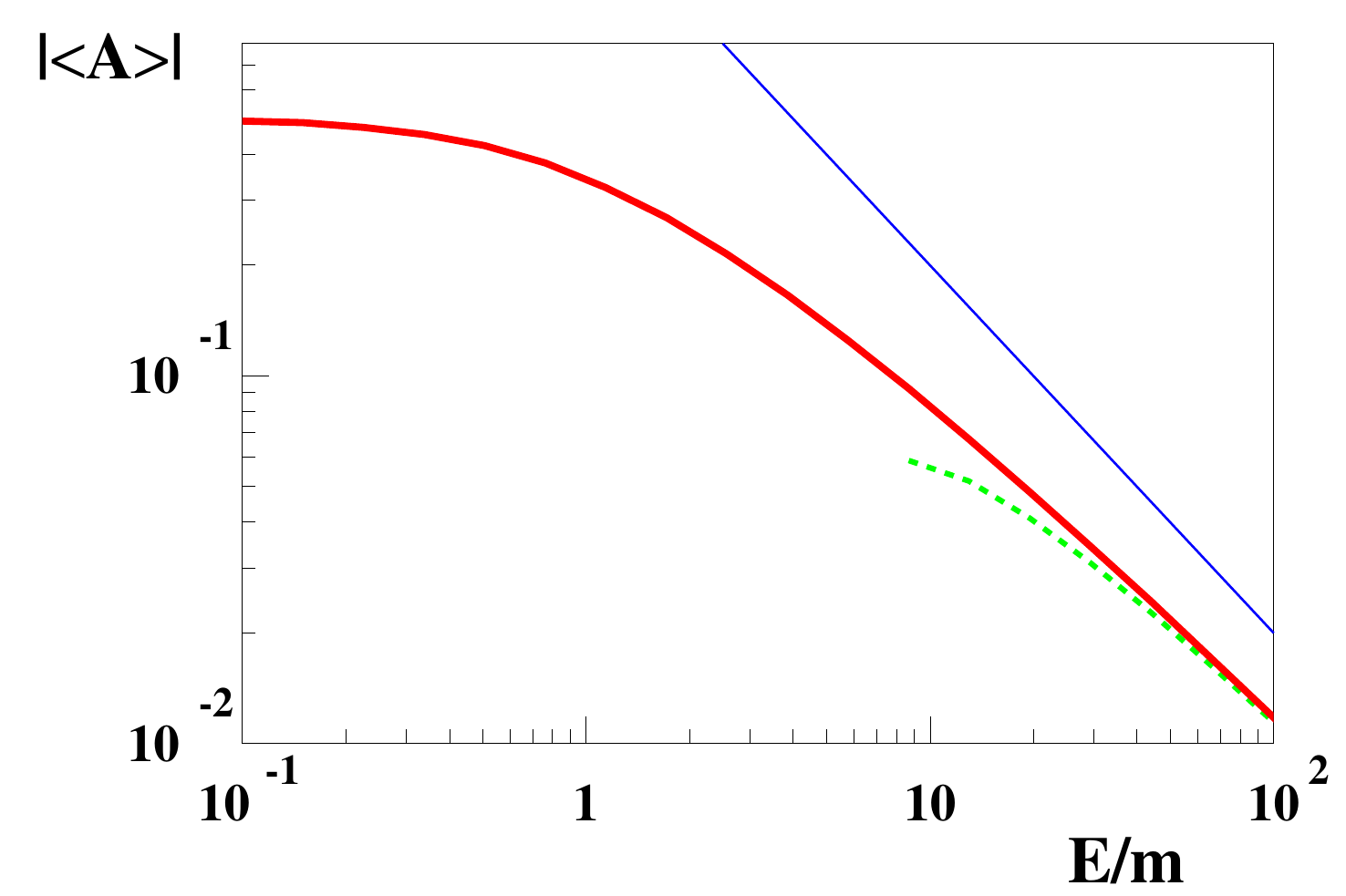}
 \caption{\label{fig:compton:asymmetry} 
Absolute value of the average asymmetry in Compton scattering on free
electrons at rest, as a function of the incident photon energy in
electron rest-mass units.
Thick solid line: full expression (eq. (\ref{eq:compton:asymmetry:full})).
Dashed line: high-energy approximation (eq. (\ref{eq:compton:asymmetry:high})).
Thin solid line: high-energy asymptote.
}
 \end{center}
\end{figure}
We now examine two limiting cases:
\begin{itemize}
\item 
At low energies, $k_0 \approx 0$, eq. (\ref{eq:phi:diff:cross:section})
reduces to:
\begin{equation}
\dd \sigma
 = \gfrac{r_0^2}{2} 
\left[
1 - \cos{(2\phi)} P /2
\right]
\frac{k_0}{3}
 \dd \phi,
\end{equation}
which results in a total cross section of $\sigma = 8 r_0^2 \pi /3$, 
i.e., the Thomson cross section.
The low-energy average asymmetry is $\langle {\cal A} \rangle = -1/2$.
\item 
At high energies, 
\begin{equation}
\dd \sigma
 = \gfrac{r_0^2}{2 k_0} 
\left[
\log(2 k_0) + \frac{1}{2} + P\frac{4-2 \log(2 k_0)}{k_0} \cos{(2\phi)}
\right]
 \dd \phi, 
\end{equation}
which results in a total cross section of $\sigma = \pi r_0^2 (\log(2 k_0) + \frac{1}{2})/k_0$.
The high-energy average asymmetry is
\begin{equation}
\langle {\cal A} \rangle = - \frac{4(\log{2 k_0}-2)}{k_0 (2\log{2 k_0}+1)}.
\label{eq:compton:asymmetry:high}
\end{equation}
\end{itemize}

The average asymmetry decreases at high energies, asymptotically approaching 
$\langle {\cal A} \rangle \approx - 2/ k_0$.
The variation of the average polarisation asymmetry of photon Compton
scattering on free electrons at rest (eq. (\ref{eq:compton:asymmetry:full}))
is compared to its high-energy approximation 
(eq. (\ref{eq:compton:asymmetry:high}))
in Fig. \ref{fig:compton:asymmetry}.
The absence of sensitivity of Compton polarimeters at high energies
\cite{McConnell:2004} is due to this strong decrease of $|\langle {\cal A} \rangle|$.

\section{Optimal variable for polarisation measurement }

The value of the polarisation fraction $P$ is classically obtained by
a fit to the $\phi$ distribution.
A way to improve the polarisation sensitivity is to make an optimal
use of the information contained in the multi-dimensional probability
density function (pdf) through the use of an optimal variable
(\cite{Bernard:2013jea} and references therein), that is, of a weight
$w(\Omega)$ such that the $P$ dependence of the expectation value
$\text{E}(w)$ of $w$ allows a measurement of $P$, and that the
variance of such a measurement is minimal.
The solution, up to a multiplicative factor, is (eg. \cite{Tkachov:2000xq}):
\begin{equation}
w_{\text{opt}} = \frac{\partial \ln p(\Omega)}{\partial P}.
\end{equation}
In the present case of a polarisation measurement:
\begin{equation}
p(\Omega) \equiv f(\Omega) + P \times g(\Omega), 
\label{eq:p(omega)f+Pg}
\end{equation}
with $\int f(\Omega) \dd \Omega = 1$ and $\int g(\Omega) \dd \Omega = 0$,
we obtain:
\begin{equation}
w_{\text{opt}} = \frac{g(\Omega)}{f(\Omega) + P \times g(\Omega)}, 
\label{eq:w:opt}
\end{equation}
that is, if $|P \times g(\Omega)|$ is small compared to $f(\Omega)$, 
\begin{equation}
w_{\text{opt}} = \frac{g(\Omega)}{f(\Omega)}. 
\label{eq:w:opt:approx}
\end{equation}

The $1^\text{st}$ moment of $w_{\text{opt}}$ is 
$ \displaystyle \langle w_{\text{opt}} \rangle = 
\int \gfrac{g(\Omega)}{f(\Omega)} \times \left[ f(\Omega) + P \times g(\Omega) \right] \dd \Omega = 
P \int \gfrac{g^2(\Omega)}{f(\Omega)} \dd \Omega $, 
which is proportional to $P$.
The expressions for $f$ and $g$ are obtained from the measured values
of $\phi$ and $\theta$ (and therefore of $x$) by equating the constant
term and the term proportional to $\cos{(2\phi)}P$ in eqs.
 (\ref{eq:p(omega)f+Pg})
 and (\ref{eq:def:diff:cross:section:compton}).
The spectrum of $w_{\text{opt}}$ is shown in
Fig. \ref{fig:compton:spectra} for a fully polarised beam.
We can see that $|w_{\text{opt}}|$ is most often much smaller than
unity (beware the vertical log scale), so that our neglecting
$|P \times g(\Omega)|$ in the expression of 
$f(\Omega) + P \times g(\Omega)$
was legitimate.
The asymmetry, the non-evenness of the $w_{\text{opt}}$ distribution
makes the non-zero average due to the beam polarisation explicit.

Moment's methods are equivalent to a likelihood analysis in the case
where the pdf is a linear function of the variables that one aims to
measure, as is the case here, but they are much simpler to instantiate
as one just has to compute $w_{\text{opt}} (\theta, \phi, x)$, and 
average it over the whole statistics.
Although the analysis of experimental data is beyond the scope of this
paper, the following considerations apply:
\begin{itemize}
\item
Background subtraction reduces to a simple subtraction in computing the average
of $w_{\text{opt}}$. Their $n$-dimensional parametrization is not needed.
\item
Likelihood methods need the use of a $n$-dimensional parametrization of the acceptance, or efficiency, for correction.
This is pretty simple in the case of Compton scattering for which 
the final state is described by only two variables, but for higher-dimensional systems, 
producing enough Monte Carlo (MC) statistics and determining a parametrization becomes a nightmare:
in that case the use of a moments-based efficiency correction becomes mandatory
(for a real-case presentation see eg., 
section IV.A, eqs. (18)-(24) and VI.B eqs. (47)-(49) of \cite{Aubert:2004cp}).
\end{itemize}

In the ``reduced'' 1D case of eq. (\ref{eq:def:diff:cross:section:compton:int}),
 $w_{\text{opt}} $ becomes $w = 2 \cos{2\phi}$ and the estimator for 
$\langle {\cal A} \rangle P$ is $\langle w \rangle$ \cite{Bernard:2013jea}.
The uncertainty then reads: 
\begin{eqnarray}
\sigma_{P} = \frac{1}{\langle {\cal A} \rangle \sqrt{N}} \sqrt{2 - (\langle {\cal A} \rangle P)^{2}}, 
 \label{eq:P:sig}
\end{eqnarray} 
that is, in the case of Thomson scattering ($\langle {\cal A} \rangle =-1/2$), 
$ \sigma_{P} \approx \sqrt{(8 - P^2)/N} $.
Needless to say, in the case where the direction of the polarisation of the
emission of a particular cosmic source ``on the sky'' is unknown, a
combined use of
$\langle 2 \cos{2\phi} \rangle$ 
and of 
$\langle 2 \sin{2\phi} \rangle$ 
should be used.

\begin{figure} [ht]
 \begin{center}
 \includegraphics[width=0.6\linewidth]{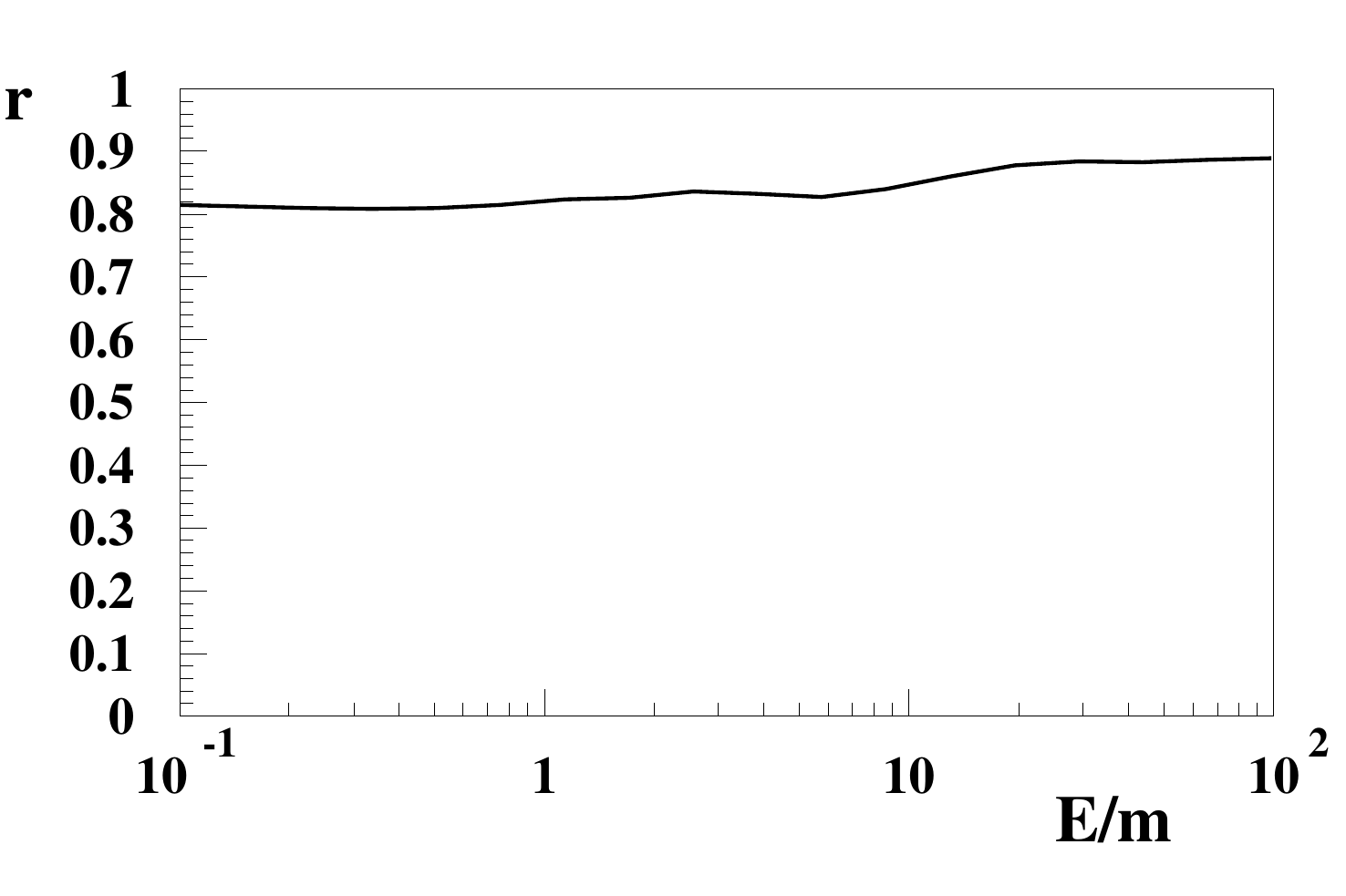}
 \caption{\label{fig:figure:of:merit:ratio} 
Ratio $r$ of the figures of merit of the 2D to 1D estimators of the
linear polarisation fraction, as a function of the incident photon
energy in electron rest-mass units.}
 \end{center}
\end{figure}
The performance of the 2D estimator $\langle w_{\text{opt}} \rangle$
is compared to that of the 1D $\langle w \rangle$ by
the comparison of the ratios of the RMS width normalized to the mean value:
\begin{equation}
r = 
 \frac{\text{RMS}_{w\text{opt}}} {\langle w_\text{opt} \rangle}
 \bigg/ 
 \frac{\text{RMS}_w} {\langle w \rangle}.
\end{equation}

In contrast with polarimetry performed with $e^+e^-$ telescopes, for
which an improvement in the precision of the measurement of the linear
polarisation fraction by a factor of larger than two is at hand
(Fig. 21 right of \cite{Bernard:2013jea}), 
 in the case of Compton polarimeters the improvement is
found to be marginal, varying from 
$\approx 20\%$ at low energy to
$\approx 10\%$ at high energy
(Fig. \ref{fig:figure:of:merit:ratio}).
These results are in qualitative agreement with those obtained at 100 keV by 
a likelihood analysis of the doubly differential cross section 
\cite{Krawczynski:2011fm}.

In summary, we have obtained the expression for the average
polarisation asymmetry, or modulation factor, of Compton scattering on 
 free electrons at rest, 
eq. (\ref{eq:compton:asymmetry:full}), 
Fig. \ref{fig:compton:asymmetry}.
We have then obtained a simple optimal estimator of the polarisation
fraction $P$ that makes use of all the information (azimuthal and polar
angles of the scatter), avoiding the technicalities of a maximum
likelihood analysis but with the same performance.

\section{Acknowledgments}

It a pleasure to acknowledge
the support by the French National Research Agency
(ANR-13-BS05-0002) and
the scrutiny and the suggestions of referee \#1 of 
Nuclear Instruments and Methods in Physics Research A.

\end{document}